\documentclass[12pt]{article}
\begin{document}
\baselineskip 3.9ex
\def\be{\begin{equation}}
\def\ee{\end{equation}}
\def\ba{\begin{array}{l}}
\def\ea{\end{array}}
\def\bea{\begin{eqnarray}}
\def\eea{\end{eqnarray}}
\def\no#1{{\tt  hep-th#1}}
\def\nn{\nonumber\\}
\def\nl{\hfill\break}
\def\ni{\noindent}
\def\nu{\noindent\underbar}
\def\bibi{\bibitem}
\def\c#1{{\hat{#1}}}
\def\eq#1{(\ref{#1})}
\def\pgap{\vspace{1.5ex}}
\def\ggap{\vspace{10ex}}
\def\gap{\vspace{3ex}}
\def\del{\partial}
\def\o{{\cal O}}
\def\z{{\vec z}}
\def\re#1{{\bf #1}}
\def\av#1{{\langle  #1 \rangle}}
\def\S{{\cal S}}
\def\sbh{S_{\rm BH}}
\def\hst{H_{\hbox{st}}}
\def\proj#1#2{| #1 \rangle \langle #2 |}
\def\tl{\tilde l_s}
\def\tg{\tilde g_s}
\def\eps{{\cal \epsilon}_0}

\renewcommand\arraystretch{1.5}
\begin{flushright}
TIFR-TH-00/61\\
hep-th/0011094
\end{flushright}
\begin{center}
\vspace{2 ex}
{\large{\bf Matrix Model, Noncommutative Gauge Theory and 
the Tachyon Potential}}\\
\vspace{10 ex}
Gautam Mandal and Spenta R. Wadia\\
\vspace{1ex}
{\sl Department of Theoretical Physics,}\\
{\sl Tata Institute of Fundamental Research,}\\
{\sl Homi Bhabha Road, Mumbai 400 005, INDIA. }\\
\vspace{1ex}
{email: mandal,wadia@theory.tifr.res.in}\\
\vspace{10 ex}
\pretolerance=1000000
\bf ABSTRACT\\
\end{center}
\vspace{1 ex} The $D2$ brane-anti-$D2$brane system is described in the
framework of BFSS Matrix model and noncommutative (NC) gauge theory.  The
physical spectrum of fields is found by appropriate gauge fixing. 
The exact tachyon potential is computed in terms of these variables
and an exact description of tachyon condensation provided.  We exhibit
multiple vortex production with increasing topological charge and
interpret this as gradual conversion of the brane-antibrane system to
$D0$ branes.  The entire analysis is carried out using the known
hamiltonian of the Matrix model, which is equivalent to the
hamiltonian of the NC gauge theory.  We identify the supersymmetric
ground state of this hamiltonian with the tachyonic vacuum; Sen's
conjecture about the latter follows simply from this identification.
We also find two types of closed string excitations, solitonic
({\it a la} Dijkgraaf, Verlinde and Verlinde) as well
as perturbative, around the tachyonic vacuum.
\vfill
\clearpage

\section{\bf Introduction}

Over the last couple of years open string theory on unstable D-branes
has been studied extensively to describe (a) lower dimensional
D-branes as solitons \cite{9902105}-\cite{man-rey}, (b) the closed
string vacuum as the tachyonic vacuum
\footnote{The phrase ``tachyonic vacuum'' refers to the global minimum
of the tachyon potential; there are no open string tachyonic
directions in this vacuum.} \cite{Kostelecky:1990nt}-\cite{tathagata}
and also (c) closed string excitations \cite{Yi:1999hd}-\cite{cse-sen}
around the tachyonic vacuum. Although the initial discussions were in
the framework of first-quantized or second-quantized open string
theories, a number of recent papers have used the framework of
noncommutative gauge theories 
\cite{Connes:1998cr}-\cite{dha-kit} which are known
\cite{Connes:1998cr,Seiberg:1999vs} to describe open strings on a
D-brane in the presence of a constant $B_{\rm NS}$ or a constant
magnetic field. Solitonic $D$-branes in this language are discussed in
\cite{DMR}-\cite{man-rey}, the tachyonic vacuum in
\cite{vacuum-gms,vacuum-sen,seiberg,tatar,shenker,li}, 
whereas closed string
excitations around the tachyonic vacuum in the NC framework are
discussed in \cite{HKLM,larsen}.

Unstable D-brane configurations which involve the ``wrong'' $Dp$-brane
($p$ odd for type IIA, even for IIB) are characterised by a {\em real}
tachyon. In the NC gauge theory or its equivalent Matrix model
formulation, such theories are mostly studied using a DBI-like action
\cite{Sen:1999md,Sen:1999xm} (or certain truncations from open string
field theories) that assumes a certain dependence on the tachyon and
gauge and matter fields on the brane. At the moment there is no first
principles derivation of this action. Although there is considerable
support for the basic features of this action, there are subtleties
regarding the choice of variables and the form of the action
especially around the tachyonic vacuum \cite{vacuum-sen}.

The situation with a brane-antibrane pair is somewhat different. Here
the tachyon is complex, which can, once again, decay \cite{9902105} to
lower dimensional branes (solitons) or the vacuum (global minimum of
the potential).  The important difference from the case of the real
tachyon, from the point of view of NC gauge theory or its equivalent
Matrix formulation, is that this system is directly describable
\cite{BSS,d2-d2b-1,d2-d2b-2} using the well-known action/hamiltonian
of the BFSS Matrix model \cite{BFSS} around a known background. This
gives us a well-defined hamiltonian framework to discuss various
features of this theory including the tachyonic vacuum.  Since
M-theory, and therefore Matrix model, in principle contains all the
states of string theory, the brane-antibrane system offers an
excellent testing ground to compare with the derivation of the same
states, including the vacuum, in open string field theory.

In this paper, we will address the issue of tachyon condensation on a
brane-antibrane system in the framework of noncommutative gauge
theory.  In Sections 2 and 3, we consider the complex tachyon of the
$D2-\overline{D2}$ system in the context of the BFSS Matrix model
using the remarkable connection of the matrix model with
noncommutative gauge theory.  In this framework the $D2-\overline{D2}$
system appears as a classical solution of the noncommutative
$U_{\infty}(2)$
\footnote{We will use the notation $U_{\infty}(2)$ to imply $U(\infty)
\otimes U(2)$ which is the gauge group of $U(2)$ noncommutative gauge
theory.} gauge theory in $2+1$ dimensions. We discuss the Hamiltonian
formulation of this theory in a unitary gauge to identify the
$U_{\infty}(1)\times U_{\infty}(1)$ gauge fields and the tachyons. We
present an exact expression for the potential energy and present an
explicit vacuum solution. In Section 4, we further gauge fix the
$U_{\infty}(1)\times U_{\infty}(1)$ symmetry to exhibit the physical
degrees of freedom: a complex tachyon field and two hermitian fields
(one each for the two photons). We exhibit the potential in terms of
the physical variables. In Section 5, we describe a sequence of
solutions in which an increasingly large number of vortices are
produced on the brane-antibrane pair to partly reduce the energy of
the initial unstable system. We show, by explicit computation of the
topological charge of these vortices, that these correspond to $D0$
branes; these facts are further verified by translating the solution
back to the Matrix model description.  In this way, the vacuum
consists of a large number of vortices ($D0$ branes) cancelling all
the energy of the brane-antibrane system, in conformity with Sen's
conjecture \cite{ORIGINAL,sen-main}. In Sec. 6 we find two classes of
closed string states around the tachyonic vacuum: solitonic ones which
are the analogs of the Matrix string of \cite{DVV}, and perturbative
flux tube-like solutions analogous to the ones discussed in
\cite{HKLM} for the case of the real tachyon.

While this work was in progress, we received \cite{shenker} which
overlaps with some aspects of the present paper. See also
\cite{li}.

\section {\bf Matrix model and noncommutative gauge theory}

The BFSS matrix model \cite{BFSS}   
is characterised by the following Hamiltonian%
\footnote{The velocity $\dot X^M$, in the
hamiltonian framework, should be
regarded  as $\dot X^M = 
g_s\, \Pi^M$.} (in  $2\pi l_s^2=
2\pi \alpha'= 1$ units)
\be
H =  \frac{\sqrt{2\pi}}{g_s}
\ {\rm Tr} \left(\frac{1}{2} \sum_{M=1}^9
(\dot{X}^M)^2  -
\frac{1}{4} \sum_{M,N} [X^M,X^N]^2
+\Psi^T\gamma_M[\Psi,X^M] \right),
\label{h}
\ee 
and the Gauss-law constraint  
\be 
[X^M,\dot{X}^M]+ i \{ \Psi^T, \Psi \}=0 
\label{g}
\ee
The $X^M , M=1,..,9$ are $N \times N$ matrices
($N$ representing the number of
$D0$-branes) and $g_s$ is the string
coupling constant.  

A classical configuration of
$k$ coincident $D2p$ branes is described by 
\cite{BSS}
\bea
X^a_{\rm cl} &=& x^a \otimes {\bf 1}_{k\times k}
,\; a=1,2,\ldots,2p, \nn
X^i_{\rm cl} &=& 0. \; i=2p+1,\ldots,9
\label{d2p}
\eea
where $x^a$ satisfy the  Heisenberg algebra:
\be 
[x^a,x^b]=i\theta^{ab} 
\label{heisenberg}
\ee
In the  case of the $D2$ brane, we will 
take 
\be \theta^{ab}= \theta \epsilon^{ab},
\, \epsilon^{12}=1
\label{d2}
\ee 
so that  \eq{heisenberg}
simply becomes
\be
[x^1, x^2] = i \theta
\label{heisen}
\ee The Heisenberg algebra \eq{heisen} can be represented in a
one-particle Hilbert space ${\cal H}$ where $x^1, x^2$ are identified
with the position and the momentum respectively, $\theta$ playing the
role of $\hbar$.

As is well-known, \eq{heisen} does not have finite dimensional
representations.  In the BFSS framework it is, however, useful (as we
will see in Section 5) often to regard the size of the matrices $X^M$
to be large but finite (it represents the total $D0$ charge or the
total ``longitudinal'' momentum in the M-theory direction). One
such  (approximate) $N$ dimensional 
representation  is constructed  through  $N \times N$
matrices $U$ and $V$, satisfying
\be U^N = V^N=1, \; UV = e^{2\pi i/N} VU
\ee
The matrices $U,V$ can be explicitly constructed  for any $N$.
Using these, one defines  $x^1, x^2$ by
\be
U = e^{i x^1},\; V= e^{i x^2} 
\label{approx}
\ee
These satisfy
\be
[x^1, x^2] \approx 2\pi i/N
\ee
the approximation becoming exact in the limit $N\to \infty$.  In
this way, the matrices in \eq{d2p} for $p=1$ will be $k N \times k N$, 
viewed as adjoints of $U(k) \otimes U(N) \equiv
U_N(k)$. Having said this, we will now go back to infinite-dimensional
representations of \eq{heisenberg} or \eq{heisen}, represented by
operators in the Hilbert space
${\cal H}$. We will denote the group of
transformations of such operators as $U(k) \otimes U(\infty) \equiv
U_\infty(k)$.

We will parameterize the bosonic fluctuations of the $X^M$
around $\eq{d2p}$ as \cite{BSS,seiberg} 
\bea 
X^a &=& x^a \otimes {\bf 1}_{k\times k}+
\theta^{ab}{\cal A}_{b}(x)
\nn
X^j &=& \phi^j 
\label{param}
\eea
Here ${\cal A}_a$ is a $k\times k$ hermitian
matrix of hermitian operators ${\cal
A}_{a,mn}(x), n,m=1,
\ldots,k$ (similarly for $\phi^j$). The fundamental observation,
made in early Matrix model literature 
\cite{BSS,KK} and
emphasized recently in \cite{seiberg,dha-wad}, is that the
group of unitary transformations on the matrices gets represented as
$U_{\infty}(k)$ gauge transformations on ${\cal A}_{a,mn}(x)$:
\be
{\cal A}_a\rightarrow {\cal U}^\dagger {\cal A}_a{\cal U} + i{\cal
U}^\dagger \del_a{\cal U}
\ee
where $\del_a=-i(\theta^{-1})_{ab}[x_b,.]$ is the 
derivative (see, e.g., \cite{Alvarez-Gaume:2000dx}) operator.

The matrix model Hamiltonian (\ref{h}) and the Gauss law (\ref{g}) can
now be written as those of a $U(k)$ noncommutative gauge theory:
\bea
H &=& \frac{\sqrt{2\pi}}{g_s}\ 
{\rm Tr} \Bigg[ \frac{1}{2} G^{ab}\dot{\cal A}_a\dot{\cal A}_b +
\frac{1}{4} G^{ac}G^{bd}
({\cal F}_{ab} -(\theta^{-1})_{ab})({\cal F}_{cd} -(\theta^{-1})_{cd})
\nn
&+& \frac{1}{2} (\dot{\phi}^{i})^2
+ \frac{1}{2} G^{ab} {\cal D}_a \phi^i {\cal D}_b \phi^i
- \frac{1}{4} \sum_{ij} [\phi^i, \phi^j]^2 \Bigg]
\label{ham}\eea
\be
G^{ab} {\cal D}_a \dot{\cal A}_b 
-i [\phi^i , \dot{\phi}^i] +
\{ \Psi^T, \Psi \}=0 \label{gauss}
\ee
In the above we have defined the field strength as
\be
{\cal F}_{ab} \equiv  \del_a {\cal A}_b - \del_b {\cal A}_a
- i [{\cal A}_a, {\cal A}_b]
\ee
and covariant derivatives as
\be
{\cal D}_a = \del_a - i [ {\cal A}_a, .]
\ee
The ``open string metric'' $G^{ab}$ is defined as
\be
G^{ab} = \sum_c \theta^{ac} \theta^{bc}
\ee

In the following we will restrict ourselves to the case $k=2$
(that is, two coincident branes) and also $p=1$ 
($D2$ branes, see \eq{d2}) mostly. The resulting gauge theory will
be a $U(2)$ noncommutative gauge theory in $2+1$
dimensions. We note that in this case
\be
G^{ab} = {\rm diag}[\theta^2, \theta^2]
\ee

\vspace{3ex}

\noindent{\large\bf The Brane-Antibrane System}

\vspace{5ex}

We now consider the following static solution \cite{shenker}
to the equations of motion of $U(2)$ non-commutative gauge theory in 
$2+1$ dimensions:
\bea
{\cal A}_1 &=&  \left( \begin{array}{cc}
		0 & 0 \\
		0 & \frac{2}{\theta}x^2
		\end{array} \right),
\quad  
{\cal F}_{12}=  \left( \begin{array}{cc}
		0 & 0 \\
		0 & -\frac{2}{\theta}
		\end{array} \right),
\nn
\phi^j &=& 0 = {\cal A}_2
\label{background}
\eea
The fields are represented here as $2 \times 2$ matrices
each entry of which is an operator in the
Hilbert space ${\cal H}$.
 
The energy of this solution in the NCYM theory can be seen to be%
\footnote{We will use the notation ``Tr''
for trace over $ U(\infty)
\otimes U(2)$ indices whereas
``tr'' will denote a   trace over
only $U(\infty)$.}
\be
{\cal E} = \epsilon_0 \, 2\, {\rm tr} {\bf 1},
\quad \epsilon_0 \equiv \frac{\sqrt{2\pi}\theta^2}{2 g_s} 
\label{energy}
\ee
which is infinite. 
This is to be expected since we are considering noncompact 2-branes.
A more useful notion is the tension of
the brane-antibrane system, defined by 
\be
M= \int d^2 x {\mathcal T}_2 = {\mathcal T}_2\,
2\pi\theta \, {\rm tr}{\bf 1}  
\label{rest-mass}
\ee
where $M$ is the rest mass, related to the light-front
energy ${\cal E}= P^-$ by the formula \cite{BFSS,susskind}
\be
M^2 = 2 {\cal E} P^+
\label{light-cone}
\ee 
In order to calculate the tension, we need to regularize the
expressions appearing in \eq{energy} and \eq{rest-mass}.  Such a
regularization is provided, {\it e.g.}, 
by \eq{approx} in which the $U_\infty(2)$
representations are regarded as $2N\times 2N$ matrices with $N$
sufficiently large. The light-front energy and rest mass then become
\be
{\cal E} = \epsilon_0\, 2 N
\label{energy-n}
\ee
\be
M ={\mathcal T}_2\, 2\pi\theta N
\ee
In our notation ${\rm tr}{\bf 1}= N, {\rm Tr} {\bf 1}= 2N$.
Using the fact that for $2N\times 2N$ matrices, the
light-cone momentum is
\be
P^+ = \frac{2N}{R_{10}}, \ R_{10}= \frac{g_s}{\sqrt{2\pi}},
\ee
we get, from \eq{light-cone}
\be
{\mathcal T}_2 = 2 {\mathcal T}_{D2},
\quad {\mathcal T}_{D2} = \frac{1}{\sqrt{2\pi}g_s}
= \frac{1}{(2\pi)^2 g_s l_s^3}
\label{tension}
\ee
which is exactly the result expected from string theory
\cite{sen-main}. In the final expression for the
D2-brane tension ${\mathcal T}_{D2}$, we have reinstated
$l_s$.

The solution \eq{background}  describes a  coincident 
$D2-\overline{D2}$ pair. It is easy to
check, using  \eq{param}, that the matrix model configuration
equivalent to \eq{background} is
\be X^{1}=x^{1}\otimes {\bf 1}, X^{2}=x^2 
\otimes \sigma_3 
\label{D22}
\ee
This is well-known \cite{d2-d2b-1,d2-d2b-2} to describe 
a coincident $D2-\overline{D2}$ pair.

\section{\bf Unitary Gauge Fixing and the Tachyon Fields}

In the discussion of tachyon condensation in the following three
sections, we will work in the ``Higgs branch'' $\phi^j=0$.  We will
relax this condition when we consider excitations around the tachyonic
vacuum.

In this section, we discuss the procedure of gauge fixing in order
to identify the physical degrees of freedom. We begin with a 
suggestive \emph{parameterization} of the $U(2)$
variables which is useful:
\be {\cal A}_{a}(x)=\left( \begin{array}{cc}
A_{a}(x) & T_{a}(x)\\
T_{a}^{\dagger }(x) & \tilde{A}_{a}(x)
\end{array} \right) \ee
The field strength is then given by:
\be {\cal F}_{ab}=
\left( \begin{array}{cc}
F_{ab}-i(T_{a}T^{\dagger }_{b}-T_{b}T^{\dagger }_{a}) 
& D_{a}T_{b}-D_{b}T_{a}\\
(D_{a}T_{b})^{\dagger }-(D_{b}T_{a})^{\dagger } 
& \tilde{F}_{ab}-i(T^{\dagger }_{a}T_{b}-T^{\dagger }_{b}T_{a})
\end{array} \right) \ee
where $F_{12}={\del_1}A_2-{\del_2}A_1 - i[A_1,A_2],
\tilde F_{12}={\del_1}
\tilde A_2-{\del_2}\tilde A_1 - i[\tilde A_1,
\tilde A_2]$ and $D_a
T_b=\del_aT_b -iA_a T_b +i T_a \tilde{A_b}$.

We now discuss the unitary gauge that fixes the $U_{\infty}(2)$ gauge 
symmetry to $U_{\infty}(1)\times U_{\infty}(1)$. The unitary
gauge, applied to the $U(2)$ field strength, says  that
the gauge field strength ${\cal F}_{12}$ is diagonal
in the $U(2)$ space.  This, in our parameterization, becomes 
\be D_{a}T_{b}=D_{b}T_{a}
\label{GF}
\ee 
and its hermitian conjugate. It
can be shown that every field configuration is gauge equivalent to one
that satisfies the gauge condition
\eq{GF} \footnote{One can see this easily
for large $2N\times 2N$ matrices.}, and further the residual
$U_{\infty}(1)\times U_{\infty}(1)$ gauge transformations are given
by:
\bea
A_a \rightarrow U^\dagger A_a U + i U^\dagger \del_a U\\
{\tilde A}_a \rightarrow \tilde U^\dagger A_a 
\tilde U + i \tilde U^\dagger\del_a \tilde U\\
T_a\rightarrow U^\dagger T_a \tilde U
\label{residual}
\eea
where $U, \tilde U$ are unitary operators on the noncommutative space
($U$ here not to be confused with the $U$ of \eq{approx}).
These transformations enable us to interpret $T_a$ as a matter field.
While there appear to be two complex matter fields, 
the gauge fixing condition
(\ref{GF}) relates them (we will detail the
solution of the gauge fixing condition later on).
The static part of the Hamiltonian (\ref{ham}),
in the unitary gauge, is given by
\bea 
\hst &=& {\cal \epsilon}_0 \,
{\rm tr}\, 
(h_1^2 + h_2^2)
\nn
h_1 &=& \theta \big[ F_{12}-
i(T_1 T^\dagger_2 - T_{2} T^\dagger_1) \big] + 
{\bf 1}
\nn
h_2 &=& \theta \big[ \tilde{F}_{12}-
i(T^\dagger_1 T_2 -T_2^\dagger T_1)
\big] +  {\bf 1}
\label{pot}
\eea
The symbol $\eps$ has been defined in \eq{energy}.
We will define the ``potential'' to be
\be
V = \hst - {\cal E} = \eps\, \left[ {\rm tr} \left(
h_1^2 + h_2^2 \right) - 2\, {\rm tr} \, {\bf 1}  \right]
\label{def-v}
\ee
Eqn. \eq{pot} makes it obvious that the global minimum
of the potential corresponds to
\be
h_1 = h_2= 0
\label{minimum}
\ee
since the operators $h_1, h_2$ are self-adjoint. 
Thus, at the minimum
of the potential, defined by \eq{minimum}, the total energy \eq{pot}
of the static configuration, vanishes. We will shortly argue that
this minimum corresponds to the phenomenon of tachyon condensation to
the vacuum. Therefore, the
potential at the minimum  exactly cancels the initial energy 
\eq{energy}
of the brane-antibrane system, leaving a state of zero
energy. This is exactly as required by the Sen
conjecture \cite{ORIGINAL,sen-main}. We note that the existence of
a simple supersymmetric hamiltonian \eq{h},\eq{ham} 
makes it obvious that the ground state energy
of the system should vanish. We also note
that \eq{minimum} is equivalent to,
in terms of the original Matrix model variables,
\be
[X^1, X^2]=0
\label{diagonal-vacuum}
\ee
In the following
we will see that such ground states do correspond
to the phenomenon of tachyon condensation to the
``vacuum''.

It is useful to write the potential 
in the Moyal star-product notation, 
as
\bea
\hst &=& \int d^2 x\ h,
\nn
h &=& {2 \pi \theta}\, \eps \, 
\left( h_1 \star h_1 + h_2 \star h_2 \right)
\label{pot-moyal}
\eea 
where $h_1, h_2$ are given by \eq{pot}. 
Clearly at this stage the static hamiltonian is translationally invariant.

In order to obtain the Hamiltonian in terms of independent degrees of
freedom we must further fix the $U(1)\times U(1)$ gauge symmetry, and
solve the Gauss law condition (\ref{gauss}).  We reiterate that the
above formulation is as yet completely general and without reference
to any specific classical configuration of the gauge theory.

\vspace{2ex}

\nu{Explicit vacuum solutions}

\vspace{2ex}

To show that there exist states with zero energy, we
explicitly exhibit a class of such states:
\bea
A_a &=& \tilde{A}_a = 0
\nn
T_1 &=& a x^2 + b x^1, \; T_2 = c x^2 + d x^1, \; ad - bc =1
\label{explicit-vacuum}
\eea
It is straightforward to check that the above
configuration satisfies \eq{minimum}. The case $(a,b,c,d)=
(0,-1,1,0)$ is discussed in \cite{shenker}.

In Section 5, we will encounter other examples of vacuum solutions. In
fact the most general solution for $A_a,\tilde A_a$, and $T_a$ that
corresponds to the ground state can be explicitly inferred by
using \eq{diagonal-vacuum} and \eq{param} in the following
way: (a) choose a basis of the Hilbert space ${\cal H}$,
(b) take $X^1, X^2$ to be arbitrary diagonal matrices
in this basis, (c) express $x^1, x^2$ in this basis,
(d) find $A_a,\tilde A_a, T_a$ using \eq{param}.

\section{The Exact Potential}

We now discuss how to fix the $U(1)\times U(1)$ gauge symmetry.  
We can, for instance, choose the background gauge
(corresponding to the background \eq{background}):
\bea 
&~&\del_1 A_1 + \del_2 A_2 =0
\nn
&~& \del_1 \tilde A_1 - i \frac{2}{\theta}[x^2,\tilde A_1]
+ \del_2 \tilde A_2 = 0
\label{bgd-gauge}
\eea
where by $A_a, \tilde A_a$ we imply the fluctuations.
It is easy to see that such a gauge choice is
translationally invariant, that is, invariant under
$ x^a \to x^a + b^a {\bf 1} $. If we are able to
solve \eq{bgd-gauge} for the independent variables,
the resulting hamiltonian will  be
translationally invariant. 

It turns out to be difficult, however, to solve 
\eq{GF} and \eq{bgd-gauge} for the independent
variables. Let us, instead, fix the
residual   $U(1)\times U(1)$ symmetry by  imposing
the axial gauge 
\be 
A_1=0, \; \tilde A_1=\frac{2x^2}{\theta}
\label{axial-gauge}
\ee
Note that any
configuration of $A_a, \tilde A_a, T_a$ can
be brought to a configuration satisfying 
\eq{axial-gauge} by applying the gauge
transformations \eq{residual}. We will
explicitly show this for four examples:

(1) The classical configuration
corresponding to two $D2$ branes, namely
\be
A_1= A_2=0, \;
\tilde A_1= \tilde A_2=0, 
\; T_1=T_2=0
\ee
can be put in the form
\be
A_1 = A_2 = 0, \;
\tilde A_1= \frac{2x_2}{\theta},
\tilde A_2= - \frac{2x_1}{\theta}, 
\;
T_1 = T_2 =0
\label{d2-d2-axial}
\ee
by choosing 
\be
U=1, \; 
\tilde U= \exp[i\pi \frac{ (x^1)^2 + (x^2)^2 }{2
\theta}]
\label{gauge-tr}
\ee
in \eq{residual}.

(2) By using the same
transformation \eq{gauge-tr} the 
vacuum solution \eq{explicit-vacuum} can be put in the
form
\bea
A_1 &=& A_2 = 0, \;
\tilde A_1= \frac{2x_2}{\theta},
\tilde A_2= - \frac{2x_1}{\theta}, 
\nn
T_1 &=& (ax^2 + bx^1) \tilde U,\, 
T_2 = (c x^2 + d x^1) \tilde U
\eea
where $\tilde U$ is as in \eq{gauge-tr}.

(3) The $n$-vortex  solutions \eq{n-vortex} that we
will encounter later can also be recast, using
\eq{gauge-tr}, as
\bea
A_1 &=& A_2 = 0, \;
\tilde A_1= \frac{2x_2}{\theta},
\tilde A_2= - \frac{2x_1}{\theta}, 
\nn
T_1 &=& \big( \sum_{j=0}^{n-1} t_j P_j
\big) \tilde U, \, T_2 = \big( \sum_{j=0}^{n-1} \omega_j P_j
\big) \tilde U
\eea
where the $t_j, \omega_j$'s are defined by
\eq{omegas},\eq{ts},\eq{rs}. Clearly the end-point
\eq{N-vortex}, the $2N$-vortex vacuum, can also be 
put in the axial gauge \eq{axial-gauge}.

(4) The fourth example is the brane-antibrane configuration
\eq{background} which already satisfies \eq{axial-gauge}. 

The point of the above exercise was to emphasize that \eq{axial-gauge}
is only a choice of gauge, and not a choice of background. Indeed the
explicit appearance of examples (2), (3) and (4) imply that we can
describe the entire process of tachyon condensation in this gauge. If
we are able to solve \eq{axial-gauge} exactly, which we will, the
resulting hamiltonian (namely, \eq{axial-gauge-ham}) will, therefore,
correspond to an {\sl exact} hamiltonian for tachyon condensation.

Let us, therefore, proceed to solve \eq{axial-gauge},
together with \eq{GF}. Using the Moyal product notation,
we get 
\be D_1 T_2-D_2 T_1 = 0 =
\del_1 T_2+\frac{2i}{\theta}T_2 \star x^2-D_2 T_1
\label{gf2}
\ee
where
\be
D_2 T_1 = \del_2 T_1 - i A_2 \star T_1 +  i T_1 \star 
\tilde A_2.
\ee
Using the fact that
\be
\del_1 T_2+\frac{2i}{\theta}T_2 \star x^2 = 
\frac{2i}{\theta} x^2 T_2,
\ee
we can solve \eq{gf2} explicitly:
\be
T_2=\frac{\theta}{2i x^2} 
(\del_2T_1 - iA_2\ast T_1+iT_1\ast \tilde
A_2)
\label{sol-gf}
\ee
Note that the independent physical variables are
$T_1, A_2, \tilde A_2$, in other words, one 
complex matter field and two real gauge fields.
This is of course exactly what one expects from
the spectrum of open string theory on the brane-antibrane system.
In the following we will denote 
\be 
T_1=T,  A_2= A,  \tilde A_2 = \tilde A
\label{independent}
\ee
In order to make the division by $x^2$ in \eq{sol-gf}
well-defined, one can use the prescription 
$1/x^2 \to 1/(x^2 + i \epsilon)$. This agrees with
the explicit solutions for $T_2$ presented in the next section.

Substituting this solution into (\ref{pot}), we get the exact static
Hamiltonian  in terms of the independent variables
$T, A, \tilde A$:
\bea
H_{st} &=& \eps\,\int \frac{dx^1 dx^2} {2\pi\theta} 
\Bigg\{ 
	\left( \theta  \del_1 A + \frac{\theta^2}{2}
	\Big[T\ast (\frac{1}{x^2} D_2\bar T)
	+ (\frac{1}{x^2}D_2 T)\ast \bar T \Big] + 1 
	\right)^2 
\nn
&~& ~~~~~~~~~~~~~~ + 
	\left( \theta \del_1 \tilde A - \frac{\theta^2}{2}
	\Big[\bar T\ast (\frac{1}{x^2} D_2 T)
	+ (\frac{1}{x^2}D_2 \bar T)\ast T \Big] - 1 
	\right)^2
\Bigg\}
\label{axial-gauge-ham}
\eea
where the bar denotes complex conjugation.
In the above equation,
\be
D_2 T = \del_2 T - i A \star T +  i T \star 
\tilde A
\ee
and the squares are evaluated using star products. $\eps$
is defined in \eq{pot}.

The potential \eq{def-v} is given by
\bea
V &=& \eps\,\int \frac{dx^1 dx^2} {2\pi\theta} 
\Bigg\{ 
	\left( \theta  \del_1 A + \frac{\theta^2}{2}
	\Big[T\ast (\frac{1}{x^2} D_2\bar T)
	+ (\frac{1}{x^2}D_2 T)\ast \bar T \Big] + 1 
	\right)^2 -1 
\nn
&~& ~~~~~~~~~~~~~~ + 
	\left( \theta \del_1 \tilde A - \frac{\theta^2}{2}
	\Big[\bar T\ast (\frac{1}{x^2} D_2 T)
	+ (\frac{1}{x^2}D_2 \bar T)\ast T \Big] - 1 
	\right)^2 -1
\Bigg\}
\label{exact-v}
\eea
It is interesting to note that the gauge fields $A, \tilde A$
appear quadratically in \eq{axial-gauge-ham},\eq{exact-v}. 
Thus, it is possible to 
eliminate them exactly and get an effective potential in terms of $T$
alone. We will not attempt to do it in this paper.

\vspace{2ex}

\nu{Gauss' law constraint}

\vspace{2ex}

It is easy to see that time-independent bosonic configurations
automatically satisfy \eq{gauss}. Therefore
the static hamiltonian and the potential mentioned above
are consistent with the Gauss' law constraint.
In Section 6 where we
will consider time-dependent $\phi^3$, we will restrict to
configurations satisfying $[\phi^3, \dot{\phi^3}]=0$,
hence satisfying Gauss' law once again.

\section{Vortex production and brane annihilation}

In order to gain more insight into the nature of the potential
\eq{exact-v} and various static solutions, it is useful to begin with
simple classes of configurations of $T, A, \tilde A$.

\vspace{2ex}

\nu{Case 1 (simple vortex)}:

\vspace{2ex}
 
Let us consider a  configuration space defined by 
\be
T =  t P_0, \; A= a P_0, \;
\tilde A = \tilde a P_0, \quad t \in {\bf C}, a, \tilde a \in {\bf R}
\label{single-vortex}
\ee
where $P_0 = \proj00$, described in the Moyal form as
\be
P_0 = 2 \exp\big[ - \frac{ r^2}{\theta} \big],
\quad r^2= (x^1)^2 + (x^2)^2
\ee
If we further specify 
$a= \tilde a$, the dependent variable $T_2$ in \eq{sol-gf}
becomes also proportional to $P_0$, with $T_2= i\, t P_0$, leading
to a consistent truncation of all the fields  
to the one-dimensional subspace $P_0$
of operators. The resulting configuration 
space clearly corresponds to a single localized vortex
at the origin of the $(x^1, x^2)$ plane. We will
shortly confirm this by explicitly calculating 
the vortex charge.

Let us now find out the minimum of the static hamiltonian
$\hst$ \eq{axial-gauge-ham} or the potential
\eq{exact-v} in the configuration \eq{single-vortex}.  It
turns out that calculations are easier in this example in the operator
formulation in which \eq{gf2} reads as 
\be
\{x^2,T_2\}=-[x^1,T_1] - A_2 T_1 + T_1 \tilde A_2 
\ee
The result is:
\be 
\frac{\hst(t,a)}{\eps}
=  2 \big[ N  - 4 \theta |t|^2 + 4 \theta^2 |t|^4 + 
\theta a^2
\big]
\ee
\be
\frac{V(t,a)}{\eps} = 
2 \big[ - 4 \theta |t|^2 + 4 \theta^2 |t|^4 + \theta a^2
\big]
\label{pot-single-vortex}
\ee
Here we have regularized ${\rm tr} {\bf 1} = N$ by
using an $N$-dimensional  Hilbert space, described
by, {\it e.g}, \eq{approx}. Note the negative sign
of the $|t|^2$, reflecting a tachyonic
mode. Indeed the tachyonic part of the hamiltonian
\eq{h}, including the kinetic term, turns out
to be
\be
H = 4 \eps \left(| \dot t|^2 - 2 \theta |t|^2 + 2 \theta^2 |t|^4  
\right)
\label{tach-ham}
\ee
which was obtained in \cite{d2-d2b-1} by explicit
Matrix model calculations. Equation \eq{tach-ham} also
appears in \cite{shenker} (Eq. (35)) whose $T$ is
related to our parameter $t$ by $T = 2 t \sqrt{\theta}$.
The mass $m_T^2= - 2\theta$ agrees with the string
theory result \cite{shenker} $\alpha'm_T^2= -\frac{1}{\pi b},
\, b = B_{12}$. 

The minimum of the potential is easily found by
noting that
\be
\frac{V(t,a)}{\eps} + 2
= 2 \big[ (1- 2 \theta |t|^2)^2 +  \theta a^2
\big]
\ee
The minimum value of the right hand side is zero,
which occurs at 
\be
|t|^2=\frac{1}{2 \theta}, a=0.
\label{classical-single-vortex}
\ee
Thus
\bea
\frac{V(t,a)_{\rm min}}{\eps} &=&  -2
\nn
\frac{H_{\rm st}(t,a)_{\rm min}}{\eps}
&=& 2N- 2
\label{energy-single-vortex}
\eea
This clearly shows that the subspace \eq{single-vortex}
describes formation of a single vortex and the classical
vortex configuration \eq{classical-single-vortex}
cancels 2 units out of 2N units of total energy of the
brane-antibrane configuration.

\vspace{2ex}

\nu{Case 2 (multiple vortices)}:

\vspace{2ex}

The above result can be easily generalized to an $n$-vortex
configuration space, given by
\be
T = \sum_{j=0}^{n-1} t_j P_j, \;
A = \tilde A = \sum_{j=0}^{n-1} a_j P_j
\label{n-vortex}
\ee
where 
\be
P_j = \proj{j}{j}
\ee
Note that Eq. \eq{n-vortex}, in the Moyal notation,
corresponds to the spherically symmetric
ansatz
\be
T(x^1, x^2) = \sum_{j=0}^{n-1} t_j 2(-1)^j L_j(2r^2/\theta) 
e^{-r^2/\theta}
\ee 
Once again, the choice $A=\tilde A$ ensures that
the dependent variable $T_2$ is of the
same form as $T$, namely,
\be
T_2 = \sum_{j=0}^{n-1} i\,\omega_j P_j
\ee
The gauge fixing conditions \eq{gf2} are solved by
\bea
\omega_{n-1} &=& t_{n-1}
\nn
\omega_{n-2} &=& \left( t_{n-2} - 2 t_{n-1} \right)
\nn
\ldots
\nn
\omega_1  &=& \left( t_1 - 2 (t_2 + \ldots + (-1)^{n-1} 
t_{n-1} ) \right)
\nn
\omega_0 &=&  \left( t_0 - 2 (t_1 + \ldots - (-1)^{n-1} 
t_{n-1} ) \right)
\label{omegas}
\eea

The static hamiltonian and the potential 
are straightforward to calculate. One gets
\bea
\frac{\hst}{\eps} &=& 2 \big[ \sum_{j=0}^{n-1}
(1 - \theta t_j \bar{\omega_j} - \theta  \bar{t_j} 
\omega_j)^2 + \sum_{j=0}^{n-2} \theta (a_j - a_{j+1})^2
+ \theta a_{n-1}^2 \big] + 2 (N-n)
\nn
\frac{V}{\eps} &=& \frac{\hst}{\eps}
- 2 N 
\eea
It is straightforward to write out these
expressions for small $n$, and discover
the tachyonic directions; we will not
write the explicit expressions here. 

Since the terms in the square brackets
in the  hamiltonian is a sum of squares, the minimum
possible value is zero. We explicitly find the
minimum, for real $t_j$, to be
\bea
a_j &=& 0
\nn
t_j &=& \frac{1}{\sqrt{\theta}} r_j, \; j=0, \ldots, n-1
\label{ts}
\eea
where the $r_j$ are defined by
\bea
2 r_{n-1}^2 - 1= 0
\nn
2 r_{n-2}^2 - 4 r_{n-2} r_{n-1} -1 =0
\nn
\ldots
\nn 
2 r_0^2 - 4 r_0( r_1 - r_2 + \ldots + (-1)^n r_{n-1}) -1 =0
\label{rs}
\eea
It is easy to see that real roots always exist
(the discriminants are always non-negative). For
example, 
\be
r_{n-1} = \frac{1}{\sqrt 2}, \; r_{n-2} = 
\frac{1}{\sqrt{2}} + 1,
\; {\rm etc.}
\label{rs-explicit}
\ee
The potential and the total energy, evaluated at the
minimum, is given by
\bea
\frac{V(\vec t,\vec a)_{\rm min}}{\eps} &=&  -2 n
\nn
\frac{H_{\rm st}(\vec t,
\vec a)_{\rm min}}{\eps}
&=& 2N-2n
\label{energy-n-vortex}
\eea
Thus the $n$-vortex solution cancels exactly $2n$
out of the total of $2N$ units of energy of the
brane-antibrane system.

\vspace{2ex}

\nu{Topological charge of the vortex}:

\vspace{2ex}

In \cite{JMW} the topological charge of vortex
configurations for a complex tachyon (assumed
to be governed by NC abelian Higgs model) was
found by essentially looking for the
``surface term'' in the hamiltonian in the
Bogomolnyi framework. Repeating a similar
analysis here, we find that the  
topological charge of these vortex solutions
are given by 
\be
{\cal I}=  {\rm Tr} {\cal F}_{12}
\label{top-charge}
\ee 
We find that the vortex charge for the solution \eq{n-vortex},\eq{ts}
is exactly 
\be {\cal I}= 2 n.  
\ee 
The factor $2$ appears because we
have normalized the topological charge to reproduce the $D0$
charge. Recall that a vortex of the complex tachyon on a
$Dp$-$\overline{Dp}$ system gives rise to BPS $D(p-2)$ branes
\cite{9902105} and the RR charge of the latter is
supposed to arise as the
topological charge of the vortex. The latter
fact has been explicitly shown in the complex tachyon model studied
in \cite{JMW}.

\vspace{2ex}

\nu{The vacuum}:

\vspace{2ex}

From the above discussion it is clear that the
tachyonic vacuum is given by $n=N$ in
\eq{n-vortex}. To be explicit, it is given by 
\be
T = \sum_{j=0}^{N-1} t_j P_j, \;
A = \tilde A = 0 
\label{N-vortex}
\ee
where $t_j$'s are defined by \eq{ts},\eq{rs}.

\vspace{2ex}

\nu{Picture of gradual annihilation}:

\vspace{2ex}

The preceding discussion provides rather compelling evidence,
within the framework of $U(2)$ NC gauge theory, that the tachyon
condensate consisting of $2N$ vortices, exactly cancels the energy of
the brane-antibrane system. This, therefore, proves Sen's conjecture.
Furthermore, we also see a concrete mechanism, within the gauge theory,
of successively lowering the energy of the system by vortex
production. The picture here is that the brane-antibrane
pair is gradually annihilating, creating more and more
vortices (which, through their topological charge,
are identified as $D0$-branes). To see this picture concretely,
let us note that 
for the $2n$-vortex solution \eq{n-vortex},\eq{ts}
\be
\frac{1}{i\theta} [X^1, X^2]
= \left( \begin{array}{lll} 
{\bf 0}_{2n} && \\
&{\bf 1}_{N-n} &   \\
&& - {\bf 1}_{N-n}   \\
\end{array}
\right)
\label{part-ann}
\ee 
Since $[X^1, X^2]$ represents $D2$-brane charge density, and since
commuting $X^1, X^2$ represents $D0$ branes, \eq{part-ann} represents
``partial annihilation'' of the brane-antibrane system to D0-branes.
The lower two blocks represent the remaining brane-antibrane system
and the lower $2n \times 2n$ block represents the D0-branes released
by partial annihilation. Note that the  brane-antibrane
configuration \eq{background} and the tachyonic vacuum 
(cf. \eq{diagonal-vacuum}) correspond
respectively to 
\be
\frac{1}{i\theta} [X^1, X^2]_{D2\,\overline{D2}}
= \left( \begin{array}{ll} 
{\bf 1}_{N-n} &   \\
& - {\bf 1}_{N-n}   \\
\end{array}
\right),
\quad 
\frac{1}{i\theta} [X^1, X^2]_{\rm vacuum}
= \left( \begin{array}{c} 
{\bf 0}_{2N}
\end{array}
\right)
\ee 

\section{Closed string excitations around the tachyonic vacuum}

In the previous sections we have described several degenerate solutions
of the tachyonic vacuum (e.g \eq{explicit-vacuum}, \eq{N-vortex}.  To
analyze the spectrum of fluctuations around any of these, one would
normally write the action or the hamiltonian of the noncommutative
gauge theory around this vacuum upto quadratic order in the variables
$T, A, \tilde A, \phi^j$. We will find it simpler, however, to go
back to the Matrix model variables $X^a, a=1,2$ and $X^i,
i=3,4,\ldots,9$ and analyze the quadratic fluctuations in terms of
them. The tachyonic vacuum, as we have seen
\eq{diagonal-vacuum}, is characterised by 
\be
[X^1, X^2]=0, \; X^i=0 
\ee 
Clearly, there is  a diagonal basis in which this will look like  
\be 
X^1 = \Lambda^1 \equiv {\rm diag}[\lambda^1_{(1)},
\lambda^1_{(2)}, \ldots], \; X^2 = \Lambda^2 \equiv {\rm
diag}[\lambda^2_{(1)}, \lambda^2_{(2)}, \ldots],\; X^i=0
\label{vacuum-bfss}
\ee
The eigenvalues $\vec X_{nn}=(\lambda^1_{(n)},\lambda^2_{(n)},
0,\ldots,0),\; n=1,\ldots, 2N$ 
represent coordinates of the $2N\; D0$ branes in the vacuum.

We will look for a closed string in the $x^9$ direction, perpendicular
to the (annihilated) brane-antibrane system. In order to describe
closed strings with a finite energy, we will compactify the $x^3$
direction on a circle of finite radius $R_9$. Matrix theory on a circle
is described \cite{taylor} by assigning appropriate periodicity
properties to the matrices $X^M$.
The hamiltonian \eq{h} becomes (after dropping the fermions) 
\be 
H = \frac{1}{g_s l_s} \int_0^{2\pi} 
\!\! \frac{d\sigma}{2\pi}\ {\rm Tr} \left(
\frac{R_9^2}{2}(\dot A)^2 + \frac{1}{2} \sum_{i=1}^8\Bigg[
(\dot {Y^i})^2 + \frac{1}{(2\pi l_s^2)^2}
\Big( R_9^2\, 
(D Y^i)^2  - \frac{1}{2}
\sum_{j} [Y^i,Y^j]^2 \Big)\Bigg]
\right)
\label{h-dual}
\ee
which can be  obtained from the 
BFSS hamiltonian \eq{h} by (i) making the
replacements \cite{taylor}
\be
X^9 \to R_9 \,iD \equiv 
R_9 \big( i\del_\sigma + A(\sigma) \big),
\quad \dot {X^3} \to R_9\, \dot A(\sigma),
\quad X^i \to  Y^i(\sigma), i=1,2,\ldots 8
\label{prescription}
\ee
and (ii) reinstating $l_s$.
It is useful to recast this in another form. Let
us (a) substitute in \eq{h-dual} the velocities $\dot A, \dot Y^i$
by their conjugate momenta, defined by 
\be
E=\frac{R_9^2}{g_sl_s} \dot A, \Pi_i = \frac{1}{g_sl_s}
\dot Y^i 
\ee
(b)  replace  $g_s, l_s$ by the M-theory variables $R_{10},
l_P$ (we denote the M-theory coordinates as
$x^0, \ldots, x^9, x^{10}$) using the relation
\be
R_{10} = g_s l_s, l_P = g_s^{1/3} l_s; 
\quad g_s = (R_{10}/l_P)^{3/2}, l_s= l_P^{3/2} R_{10}^{-1/2}
\label{m2a}
\ee
and (c) rescale
\be
Y^i \to \big( \frac{2\pi l_P^3}{R_9} \big)^{1/2},
\Pi_i \to \big( \frac{2\pi l_P^3}{R_9} \big)^{-1/2} \Pi_i
\label{rescale}
\ee
We get
\be
H =  \int_0^{2\pi} 
\!\! \frac{d\sigma}{2\pi}\ \left( {\rm Tr} \Big[
\frac{R_9R_{10}}{4\pi l_P^3} \sum_{i=1}^8\big(\Pi_i^2 +
(D Y^i)^2 \big) + \big(\frac{R_{10}}{2R_9^2} \big)^3
\big(E^2 
  - \frac{1}{2}\sum_{i,j} 
[Y^i,Y^j]^2 \big) \Big]
\right)
\label{h-m}
\ee 
The  hamiltonian \eq{h-m} can be regarded as a description of
11-dimensional M-theory on a torus of radii $(R_9,R_{10})$ (in which
$R_{10}$ denotes a light-like compactification
\cite{susskind,sei-dlcq,sen-dlcq} of M-theory, with
states characterized by 
the light-like momentum $P^+ = \frac{2N}{R_{10}}$).

The Gauss law constraint \eq{g}, in
the variables of \eq{h-dual}, reads 
\be
i \del_\sigma \dot A + [A, \dot A] + \frac{1}{R_9}[Y^i, \dot Y^i]
=0
\label{g2}
\ee
The vacuum of \eq{h-dual} should satisfy
\be
\dot A=0, i\del_\sigma Y^i + [A, Y^i]=0,
\dot Y^i=0, [Y^i, Y^j]=0
\label{vac-cond}
\ee
The tachyonic vacuum \eq{vacuum-bfss}, which  in the
variables of \eq{h-dual}, reads
\be
A(\sigma)=0, Y^1(\sigma)= \Lambda^1, Y^2(\sigma)= \Lambda^2,
Y^i(\sigma)=0, i=3,\ldots,8,
\label{tach-vac-dual}
\ee
clearly satisfies \eq{vac-cond}.

We will consider now fluctuations of \eq{h-dual} or \eq{h-m} 
around the tachyonic vacuum \eq{tach-vac-dual}. We will
consider two special regions of the parameter
space $(R_9, R_{10})$:\\
Region A:  $R_{10} \ll l_P, R_{9} \sim l_P$\\
Region B:  $R_9 \ll l_P, R_{10} \sim l_P$,\\
Using \eq{m2a} it is clear that the original
weakly coupled NC gauge theory refers to Region A.
We will however find it more convenient to
consider region B first.

\vspace{2ex}

\nu{Region B}:

\vspace{2ex}

The analysis of \eq{h-m} in this region is exactly
the same as in \cite{DVV}. We will, therefore,
mention only the salient points. In this
region it is more appropriate to obtain a type IIA
theory by compactifying M-theory on $x^9$
(rather than on $x^{10}$).
The string length $\tl$ and string coupling
$\tg$ are given by 
\be
R_{9} = \tg \tl, l_P = \tg^{1/3} l\tl; 
\quad \tg = (R_{9}/l_P)^{3/2}, \tl= l_P^{3/2} R_{9}^{-1/2}
\label{m2at}
\ee
Clearly $\tg \ll 1$. As shown by \cite{DVV},
in this limit, the off-diagonal fluctuations of
$Y^i$ are heavy, so that $[Y^i, Y^j]=0$. The
fluctuations of $E$ are also heavy. The lightest sector
is described by  a $1+1$ dimensional conformal
field theory on $S^{2N}(R^8)$. The states of
this theory are a collection of weakly coupled closed strings
(coupling $\tg$) of various lengths $n_i$ with $\sum_i n_i=
2N$. The mass scale of these strings is
\be
\tilde m_s = \frac{1}{\tl} = \frac{R_9^{1/2}}{l_P^{3/2}}
\label{mass-dvv}
\ee 
As mentioned in \cite{DVV}, these closed string states are BPS
states. Therefore these states exist for all values of the coupling
and we can continue the mass formula beyond Region B, in particular
to Region A.  In this region, the weakly coupled IIA theory is
obtained by original compactification on $x^{10}$, described by
\eq{m2a}. \eq{mass-dvv} becomes
\be
\tilde m_s = \frac{R_9^{1/2}}{g_s^{1/2}l_s^{3/2}}
\ee
which is clearly a massive state for  small $g_s$. 
These are therefore solitonic states in the
theory with small $g_s$. 

It is straightforward to see that the tachyonic vacuum
\eq{tach-vac-dual} corresponds to the ground state of the the
superconformal field theory on  $S^{2N}(R^8)$.

{\em This, therefore, establishes that there are solitonic 
closed string states around  the tachyonic vacuum.}

\vspace{2ex}

\nu{Region A}:

\vspace{2ex}

As we remarked already, this region corresponds to
the original coupling $g_s$ being weak. In this
region the closed string states constructed
above using the transverse coordinates $Y^i$ are heavy.
We will find light states here which correspond to
electric flux lines along $x^9$. These are similar to the
ones found for unstable $Dp$ branes.

We begin, therefore, by considering fluctuations of
$A(\sigma)$ in \eq{h-dual}
around the tachyonic vacuum \eq{tach-vac-dual}.
For a generic choice of the eigenvalues of
$\Lambda^1, \Lambda^2$ the $U(2N)$ symmetry will
be completely broken to $U(1)$ factors. In this
case, the $[A, Y^1]^2 + [A, Y^2]^2$ term will
give mass to the off-diagonal fluctuations of $A$.
Since we are looking for the lightest states,
we will therefore restrict to $A$
of the diagonal form
\be
A  = \Lambda = {\rm diag}[
\lambda_1, \lambda_2, \ldots, \lambda_{2N}]
\ee
For such a gauge field, the Gauss law
reads (in the absence of
fluctuations of the $Y^i$'s)
\be
\del_\sigma \dot \Lambda=0
\ee

With the above ingredients, the hamiltonian
\eq{h-dual} becomes 
\be
H = \frac{R_9^2}{2 g_s l_s} 
\sum_{i=1}^{2N} (\dot \lambda_i)^2 = 
\frac{g_s l_s}{2R_9^2} \sum_{i=1}^{2N} e_i^2 
\label{h-free}
\ee
where the canonical momenta (electric field
eigenvalues) $e_i$ are
given by
\be
e_i= \frac{R_9^2}{g_sl_s} \dot \lambda_i
\ee
 
We will show in a simple example that the above spectrum
of electric fluxes describes closed
string states in the dual type IIB description. Recall that the BFSS
hamiltonian \eq{h-dual} describes a system of $2N$ 
$D$-strings in a dual type IIB theory described by a radius $R'_9$
and coupling constant $g'_s$
\be
R'_9 = \frac{l_s^2}{R_9}, \quad  g'_s = g_s \frac{l_s}{R_9}
\ee
Consider a bound state of $n$ fundamental strings and a single
$D$-string (say the first one). This 
is described by an electric flux $e_1$
on the world-volume of the first $D$-string
(and no flux on the others), satifying
\be
e_1= n, e_i= 0, i=2,\ldots 2N
\label{fluxes}
\ee
The energy of such a configuration, according to
\eq{h-free}, is
\be
{\cal E} = \frac{g_sl_s}{2R_9^2} n^2
= g'_s \frac{R'_9}{2l_s^2}  n^2
\label{h-string}
\ee
From the above (light-front) energy we can
get the rest mass $M$ by using \eq{light-cone} with $P^+=1/R$
(corresponding to a single $D$ string carrying the
flux). The result is 
\be
M = n \frac{2\pi R'_9}{2\pi l_s^2}
\label{iib}
\ee
which is exactly the mass of a IIB string of winding number
$n$. 
 
This proves that the state we have constructed above is a 
closed string state. In the type IIB description, these are winding
states, whereas in the type IIA description these will correspond to
Kaluza-Klein states. For an $(n,k)$ bound state one needs
to take into account the permutation group $S(k)$ 
which played a role in our discussion of the solitonic
strings. Our result above can be
generalized to this case by arguments similar to the ones used
in \cite{DVV} to describe bound states of $D0$ branes
and Matrix strings. 

We note that if $\Lambda^{1,2}$ are taken to be diagonal in the simple
harmonic oscillator basis, the flux configuration \eq{fluxes}
corresponds to a localized functions (given by Laguerre polynomials)
in the $(x^1, x^2)$ plane. This corresponds to a closed flux tube
(along $x^9$) having a finite width in the $(x^1, x^2)$ directions.

\section{Discussion}

We have shown in this paper that tachyon condensation can be described
in the context of brane-antibrane pairs in a well-defined Hamiltonian
framework \eq{h} with a supersymmetric ground state. The existence of
such a Hamiltonian framework guarantees that Sen's conjecture must be
true, as we have found out explicitly. We have further shown that
there are two types of closed string excitations around the tachyonic
vacuum. The first type is solitonic (mass $\propto g_s^{-1/2}$) which
are essentially the Matrix strings of \cite{DVV}. The second type are
flux lines whose masses are small.

We conclude with a couple of remarks:

(1) It has been noted %
\cite{ORIGINAL,Yi:1999hd,Sen:1999md,Sen:1999xm,Bergman:2000xf} that
the phenomenon of tachyon condensation incorporates the Higgs
mechanism. The ``symmetric vacuum'' (top of the potential) of the
Higgs model corresponds to the perturbative vacuum of open string
theory on the brane; at the stable vacuum the gauge fields, which are
open string fluctuations, are eaten up by the Higgs mechanism.  It has
been suggested recently \cite{shatashvili} that in fact the {\em
entire tower} of open strings is ``Higgsed'', giving rise to closed
string degrees of freedom.  This viewpoint is rather interesting, and
a thorough understanding of such a claim would appear to require open
string field theory.  Since $U_\infty(k)$ NC gauge theory on branes,
equivalently the Matrix model formulation, incorporates many features
of open string field theory, it seems to provide an ideal set-up for
addressing the stringy Higgs mechanism. It will be very interesting
from this point of view to analyze the complete set of fluctuations of
the NC gauge theory described in this paper both at the top of the
potential as well as at the tachyonic vacuum.

(2) We have written down an exact tachyon potential \eq{exact-v} in
this paper, including the gauge fields.  It would be interesting to
compare our result with the various proposals for action for
brane-antibrane pairs, e.g. in \cite{Bergman:2000xf,exact-2}.  It is
also interesting to see how our action connects to the ones describing
the case of the real tachyon on unstable $Dp$-branes.

(3) The discussion in our paper has been mostly on a single
brane-antibrane pair. The generalization to $k$ brane-antibrane pairs
is an important problem. As explained in Section 2, this involves
$U(2k)$ noncommutative gauge theory. The brane-antibrane system is an
unstable, nonsupersymmetric state of this gauge theory. It is
important to study the large $k$ limit of these theories and study the
tachyon potential as well as explore possible connection with
nonsupersymmetric black holes.
 
\vspace{5ex}

\nu{Acknowledgement}: We would like to thank K.P. Yogendran for
discussions during the early stage of this work.


\begin{thebibliography}{999}

\baselineskip 0.4cm


\bibitem{9902105}
A.~Sen,
``Descent relations among bosonic D-branes,''
Int.\ J.\ Mod.\ Phys.\  {\bf A14}, 4061 (1999)
[hep-th/9902105].

\bibitem{ORIGINAL}
A.~Sen,
``Stable non-BPS bound states of BPS D-branes,''
JHEP {\bf 9808}, 010 (1998)
[hep-th/9805019]; 
``Tachyon condensation on the brane antibrane system,''
JHEP {\bf 9808}, 012 (1998)
[hep-th/9805170]; 
``BPS D-branes on non-supersymmetric cycles,''  
JHEP {\bf 12}, 021 (1998)  
hep-th/9812031.

\bibitem{sen-main}
A.~Sen,
``Non-BPS states and branes in string theory,''
hep-th/9904207.

\bibitem{9810188}
E.~Witten,
``D-branes and K-theory,''
JHEP {\bf 12}, 019 (1998)
hep-th/9810188; 
``Overview of K-theory applied to strings,''
hep-th/0007175.

\bibitem{9812135}
P.~Horava,
``Type IIA D-branes, K-theory, and matrix theory,''
Adv.\ Theor.\ Math.\ Phys.\ {\bf 2}, 1373 (1999)
hep-th/9812135.

\bibitem{RECK}
A.~Recknagel and V.~Schomerus,
``Boundary deformation theory and moduli spaces of D-branes,''
Nucl.\ Phys.\ {\bf B545}, 233 (1999)
hep-th/9811237.

\bibitem{RECK1}
C.G.~Callan, I.R.~Klebanov, A.W.~Ludwig and J.M.~Maldacena,
``Exact solution of a boundary conformal field theory,''
Nucl.\ Phys.\ {\bf B422}, 417 (1994)
hep-th/9402113.

\bibitem{RECK2}
J.~Polchinski and L.~Thorlacius,
``Free fermion representation of a boundary conformal field theory,''
Phys.\ Rev.\ {\bf D50}, 622 (1994)
hep-th/9404008.

\bibitem{Harvey:2000qu}
J.~A.~Harvey, P.~Horava and P.~Kraus,
``D-sphalerons and the topology of string configuration space,''
JHEP {\bf 0003}, 021 (2000)
[hep-th/0001143].

\bibitem{Harvey:2000tv}
J.~A.~Harvey and P.~Kraus,
``D-branes as unstable lumps in bosonic open string field theory,''
JHEP {\bf 0004}, 012 (2000)
[hep-th/0002117].

\bibitem{deMelloKoch:2000ie}
R.~de Mello Koch, A.~Jevicki, M.~Mihailescu and R.~Tatar,
``Lumps and p-branes in open string field theory,''
Phys.\ Lett.\  {\bf B482}, 249 (2000)
[hep-th/0003031].

\bibitem{Moeller:2000jy}
N.~Moeller, A.~Sen and B.~Zwiebach,
``D-branes as tachyon lumps in string field theory,''
JHEP {\bf 0008}, 039 (2000)
[hep-th/0005036].

\bibitem{deMelloKoch:2000xf}
R.~de Mello Koch and J.~P.~Rodrigues,
``Lumps in level truncated open string field theory,''
hep-th/0008053.

\bibitem{moeller}
N.~Moeller,
``Codimension two lump solutions in string field theory and tachyonic
theories,''
hep-th/0008101.


\bibitem{DMR}
K.~Dasgupta, S.~Mukhi and G.~Rajesh,
``Noncommutative tachyons,''
JHEP {\bf 0006}, 022 (2000)
[hep-th/0005006].

\bibitem{HKLM}
J.~A.~Harvey, P.~Kraus, F.~Larsen and E.~J.~Martinec,
``D-branes and strings as non-commutative solitons,''
JHEP {\bf 0007}, 042 (2000)
[hep-th/0005031].

\bibitem{wit-nct}
E.~Witten,
``Noncommutative tachyons and string field theory,''
hep-th/0006071.

\bibitem{JMW}
D.~P.~Jatkar, G.~Mandal and S.~R.~Wadia,
``Nielsen-Olesen vortices in noncommutative Abelian Higgs model,''
hep-th/0007078.

\bibitem{0008023}
Y.~Hikida, M.~Nozaki and T.~Takayanagi,
``Tachyon condensation on fuzzy sphere and noncommutative solitons,''
hep-th/0008023.

\bibitem{0008064}
J.~A.~Harvey, P.~Kraus and F.~Larsen,
``Tensionless branes and discrete gauge symmetry,''
hep-th/0008064.

\bibitem{man-rey}
G.~Mandal and S.~Rey,
``A note on D-branes of odd codimensions from noncommutative tachyons,''
hep-th/0008214.


\bibitem{Kostelecky:1990nt}
V.~A.~Kostelecky and S.~Samuel,
``On A Nonperturbative Vacuum For The Open Bosonic String,''
Nucl.\ Phys.\  {\bf B336}, 263 (1990).

\bibitem{Sen:2000nx}
A.~Sen and B.~Zwiebach,
``Tachyon condensation in string field theory,''
JHEP {\bf 0003}, 002 (2000)
[hep-th/9912249].

\bibitem{Berkovits:2000hf}
N.~Berkovits, A.~Sen and B.~Zwiebach,
``Tachyon condensation in superstring field theory,''
hep-th/0002211.

\bibitem{Moeller:2000xv}
N.~Moeller and W.~Taylor,
``Level truncation and the tachyon in open bosonic string field theory,''
Nucl.\ Phys.\  {\bf B583}, 105 (2000)
[hep-th/0002237].

\bibitem{DeSmet:2000dp}
P.~De Smet and J.~Raeymaekers,
``Level four approximation to the tachyon potential in superstring field
theory,''
JHEP {\bf 0005}, 051 (2000)
[hep-th/0003220];
``The tachyon potential in Witten's superstring field theory,''
JHEP {\bf 0008}, 020 (2000)
[hep-th/0004112].

\bibitem{Rastelli:2000iu}
L.~Rastelli and B.~Zwiebach,
``Tachyon potentials, star products and universality,''
hep-th/0006240.


\bibitem{Kostelecky:2000hz}
V.~A.~Kostelecky and R.~Potting,
``Analytical construction of a nonperturbative vacuum 
for the open  bosonic mstring,''
hep-th/0008252.

\bibitem{vacuum-gms}
R.~Gopakumar, S.~Minwalla and A.~Strominger,
``Symmetry restoration and tachyon condensation in open string theory,''
hep-th/0007226.

\bibitem{vacuum-sen} A.~Sen,
``Some issues in non-commutative tachyon condensation,''
hep-th/0009038.

\bibitem{tatar}
R.~Tatar,
``A note on non-commutative field theory and 
stability of brane-antibrane  systems,''
hep-th/0009213.


\bibitem{shenker}
P.~Kraus, A.~Rajaraman and S.~Shenker,
``Tachyon condensation in noncommutative gauge theory,''
hep-th/0010016.


\bibitem{li}
M.~Li,
``Note on noncommutative tachyon in matrix models,''
hep-th/0010058.



\bibitem{Sen:1999md}
A.~Sen,
``Supersymmetric world-volume action for non-BPS D-branes,''
JHEP {\bf 9910} (1999) 008
[hep-th/9909062].

\bibitem{Sen:1999xm}
A.~Sen,
``Universality of the tachyon potential,''
JHEP {\bf 9912} (1999) 027
[hep-th/9911116].


\bibitem{exact-1}
A.~A.~Gerasimov and S.~L.~Shatashvili,
``On exact tachyon potential in open string field theory,''
hep-th/0009103.


\bibitem{exact-2}
D.~Kutasov, M.~Marino and G.~Moore,
``Some exact results on tachyon condensation in string field theory,''
hep-th/0009148.

\bibitem{tathagata}
S.~Dasgupta and T.~Dasgupta,
``Renormalization group analysis of tachyon condensation,''
hep-th/0010247.


\bibitem{Yi:1999hd}
P.~Yi,
``Membranes from five-branes and fundamental strings from Dp branes,''
Nucl.\ Phys.\  {\bf B550}, 214 (1999)
[hep-th/9901159].

\bibitem{Bergman:2000xf}
O.~Bergman, K.~Hori and P.~Yi,
``Confinement on the brane,''
Nucl.\ Phys.\  {\bf B580}, 289 (2000)
[hep-th/0002223].

\bibitem{Gibbons:2000hf}
G.~Gibbons, K.~Hori and P.~Yi,
``String fluid from unstable D-branes,''
hep-th/0009061.

\bibitem{larsen}
F.~Larsen,
``Fundamental Strings as Noncommutative Solitons,''
hep-th/0010181.

\bibitem{cse-sen}
A.~Sen,
``Fundamental strings in open string theory at the tachyonic vacuum,''
hep-th/0010240.


\bibitem{Connes:1998cr}
A.~Connes, M.~R.~Douglas and A.~Schwarz,
``Noncommutative geometry and matrix theory: Compactification on tori,''
JHEP {\bf 9802}, 003 (1998)
[hep-th/9711162].

\bibitem{Seiberg:1999vs}
N.~Seiberg and E.~Witten,
``String theory and noncommutative geometry,''
JHEP {\bf 9909}, 032 (1999)
[hep-th/9908142].

\bibitem{Gopakumar:2000zd}
R.~Gopakumar, S.~Minwalla and A.~Strominger,
``Noncommutative solitons,''
JHEP {\bf 0005}, 020 (2000)
[hep-th/0003160].

\bibitem{Gross:2000wc}
D.~J.~Gross and N.~A.~Nekrasov,
``Monopoles and strings in noncommutative gauge theory,''
JHEP {\bf 0007} (2000) 034
[hep-th/0005204];

\bibitem{Alvarez-Gaume:2000dx}
L.~Alvarez-Gaume and S.~R.~Wadia,
``Gauge theory on a quantum phase space,''
hep-th/0006219.


\bibitem{seiberg}
N.~Seiberg,
``A note on background independence in noncommutative gauge theories,
matrix model and tachyon condensation,''
JHEP {\bf 0009}, 003 (2000)
[hep-th/0008013].


\bibitem{das-rey} 
S.~R.~Das and S.~Rey,
``Open Wilson lines in noncommutative gauge theory and 
tomography of  holographic dual supergravity,''
Nucl.\ Phys.\  {\bf B590} (2000) 453
[hep-th/0008042].


\bibitem{Gross:2000ba}
D.~J.~Gross, A.~Hashimoto and N.~Itzhaki,
``Observables of non-commutative gauge theories,''
hep-th/0008075.


\bibitem{dha-wad}
A.~Dhar and S.~R.~Wadia,
``A note on gauge invariant operators in noncommutative gauge theories  and the matrix model,''
hep-th/0008144, to appear in Phys. Lett. B.

\bibitem{dha-kit}
A.~Dhar and Y.~Kitazawa,
``Wilson loops in strongly coupled noncommutative gauge theories,''
hep-th/0010256.




\bibitem{BSS} ``Branes from Matrices''
T.~Banks, N.~Seiberg and S.~Shenker,
``Branes from matrices,''
Nucl.\ Phys.\  {\bf B490} (1997) 91
[hep-th/9612157].

\bibitem{d2-d2b-1}
O.~Aharony and M.~Berkooz,
``Membrane dynamics in M(atrix) theory,''
Nucl.\ Phys.\  {\bf B491}, 184 (1997)
[hep-th/9611215].

\bibitem{d2-d2b-2}
G.~Lifschytz and S.~D.~Mathur,
``Supersymmetry and membrane interactions in M(atrix) theory,''
Nucl.\ Phys.\  {\bf B507}, 621 (1997)
[hep-th/9612087].

\bibitem{BFSS} T.~Banks, W.~Fischler, S.~H.~Shenker and L.~Susskind,
``M theory as a matrix model: A conjecture,'' Phys.\ Rev.\ {\bf D55},
5112 (1997) [hep-th/9610043]. For reviews, see, e.g.,
\cite{banks,taylor-review}.

\bibitem{banks}
T.~Banks,
``TASI lectures on matrix theory,''
hep-th/9911068;
``Matrix theory,''
Nucl.\ Phys.\ Proc.\ Suppl.\  {\bf 67} (1998) 180
[hep-th/9710231].

\bibitem{taylor-review}
W.~I.~Taylor,
``The M(atrix) model of M-theory,''
hep-th/0002016.

\bibitem{DVV} 
R.~Dijkgraaf, E.~Verlinde and H.~Verlinde,
``Matrix string theory,''
Nucl.\ Phys.\  {\bf B500} (1997) 43
[hep-th/9703030].

\bibitem{KK}
E.~Keski-Vakkuri and P.~Kraus,
``Notes on branes in matrix theory,''
Nucl.\ Phys.\  {\bf B510} (1998) 199
[hep-th/9706196].


\bibitem{taylor} 
W.~I.~Taylor,
``D-brane field theory on compact spaces,''
Phys.\ Lett.\  {\bf B394} (1997) 283
[hep-th/9611042].

\bibitem{susskind} 
L.~Susskind,
``Another conjecture about M(atrix) theory,''
hep-th/9704080.

\bibitem{sei-dlcq}
N.~Seiberg,
``Why is the matrix model correct?,''
Phys.\ Rev.\ Lett.\  {\bf 79} (1997) 3577
[hep-th/9710009].

\bibitem{sen-dlcq}
A.~Sen,
``D0 branes on T(n) and matrix theory,''
Adv.\ Theor.\ Math.\ Phys.\  {\bf 2} (1998) 51
[hep-th/9709220].




\bibitem{shatashvili}
A.~A.~Gerasimov and S.~L.~Shatashvili,
``Stringy Higgs Mechanism and the Fate of Open Strings,''
hep-th/0011009.

\end{thebibliography}
\end{document}